# A comparison of mobile VR display running on an ordinary smartphone with standard PC display for P300-BCI stimulus presentation

G. Cattan, A. Andreev, C. Mendoza and M. Congedo

*Abstract*— A brain-computer interface (BCI) based on electroencephalography (EEG) is a promising technology for enhancing virtual reality (VR) applications—in particular, for gaming. We focus on the so-called P300-BCI, a stable and accurate BCI paradigm relying on the recognition of a positive event-related potential (ERP) occurring in the EEG about 300 ms post-stimulation. We implemented a basic version of such a BCI displayed on an ordinary and affordable smartphone-based head-mounted VR device: that is, a mobile and passive VR system (with no electronic components beyond the smartphone). The mobile phone performed the stimuli presentation, EEG synchronization (tagging) and feedback display. We compared the ERPs and the accuracy of the BCI on the VR device with a traditional BCI running on a personal computer (PC). We also evaluated the impact of subjective factors on the accuracy. The study was within-subjects, with 21 participants and one session in each modality. No significant difference in BCI accuracy was found between the PC and VR systems, although the P200 ERP was significantly wider and larger in the VR system as compared to the PC system.

## I. INTRODUCTION

Examining science fiction and fantasy literature, it appears that people enjoy stories in which characters "enter" physically into another world. *The NeverEnding Story* (Wolfgang Peterson, 1984, Germany/USA), *Jumanji* (Chris Van Allsburg, 1981, USA), *Tron* (Steven Lisberger, 1982, USA), *Narnia* (C.S. Lewis, 1950-1956, UK) and *His Dark Materials* (Philip Pullman, 1995, UK) are just a few examples where the protagonist enters another world or embodies a fictional character. In *Jumanji*, for instance, the characters become pawns in a board game, and in *Narnia*, children enter a parallel hidden world through a wardrobe, suddenly becoming warriors and princesses. This embodiment fantasy, which has been interpreted as a desire to escape reality by the personification of someone else [1], [2], partially explains the widespread interest in virtual reality (VR) technology. In fact, VR provides a means to enhance the immersion, thus reducing the distance between the user and the avatar who enters another world. Taking this a step further, incorporating brain–computer interface (BCI) technology into VR is potentially a promising step to improve the feeling of immersion. The present work contributes to the fusion of VR and BCI technology by implementing and testing a BCI displayed on VR devices running on ordinary smartphones—that is, a potentially *ubiquitous* VR technology, enabling the widespread diffusion of such technology.

A BCI is an interface that allows for direct communication between the brain and an electronic device, bypassing the usual muscular and peripheral nerve pathways [3]. Research on BCIs started in the early 1970s with the work of Vidal and collaborators [4], who designed an interface to control a cursor on a computer screen using only electroencephalography (EEG) signals [5]. Further research has strived to adapt BCI technology for people suffering from severe motor disabilities [6]–[8]. More recent is the incept of BCI technology for the general public (*e.g.,* [9]). These applications face several limitations: cumbersomeness, cost of EEG hardware, lack of reactivity of the system (low accuracy and/or low bit rate) and the need for calibration before each BCI usage [10].

For several reasons, electroencephalography (EEG) is the most suitable BCI modality to be used for the general public: it is noninvasive, transportable and inexpensive. Traditionally, BCI applications relying on EEG use three different paradigms: namely, steady-state-visually-evoked potentials (SSVEP), P300 event-related potential (ERP) and mental imagery (MI). SSVEP and P300 require sensorial stimulation of the user. They are named synchronous protocols because the interface decides when to send the stimulation, hence when the user can emit a command. By contrast, MI is defined as asynchronous, since the user may decide when to give a command by a mental imagery task such as movement imagination [3]. In the present study, we focus on P300-based BCIs. The P300 is an ERP produced by the brain about 300 ms after the presentation of a stimulus. We chose the P300 because this paradigm has a higher bit rate than MI while being less visually fatiguing than SSVEP. It can also allow the selection of items from among a large number of options, whereas SSVEP and MI are practically limited to allowing selection among just a few items [11], [12]. Moreover, [13], [14] have reported an adaptive P300 BCI that does not require calibration. This makes BCI technology more suitable for the general public, since avoiding the need for calibration is a key feature in providing a plug-and-play technology [15]. For the other main bottleneck, the encumbrance and cost of EEG hardware, the readiness of BCI technology is a matter of time, since both the bulkiness and the cost are currently being rapidly reduced (*e.g.,* OpenBCI, New York, US).

For VR, we focus on head-mounted displays (HMDs, Figure 1). The HMD we chose consists of a smartphone plugged into a plastic mask placed in front of the eyes. A software plugin, such as Google Cardboard (Google, Mountain View, US), helps to split the screen of the smartphone into two sections, each section rendering the virtual scene for a different eye. We distinguish passive HMDs, which do not incorporate any electromagnetic components, from active HMDs, which do. Active HMDs provide better interaction with the VR, thanks to the incorporated speed and proximity sensors. In fact, standard sensors in mid-range smartphones accumulate an excessive amount of drift, which results in substantial imprecision when tracking the user position and orientation. Nevertheless, in our study, we chose to focus on the use of passive HMDs because they are affordable for the general public and adapt to most currently available smartphones (Figure 1). To avoid unnecessary recalibration during the experiment, which did not involve any physical interactions, we chose to disable the smartphone sensors.

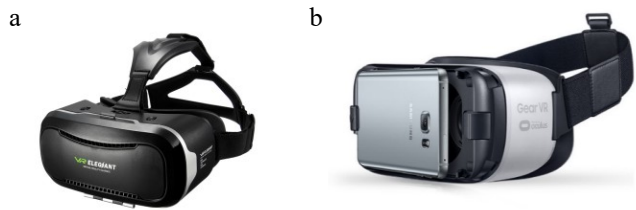

Figure 1. VRElegiant (a) and SamsungGear (b) are two popular HMDs. VRElegiant (Elegiant, Austin, US) is a passive HMD: it does not incorporate electronics. The SamsungGear (Samsung, Seoul, South Korea) is an active HMD that works only with specific smartphones from the manufacturer.

Several previous studies have focused on the integration of BCI technology with VR gaming [10], [16]–[19]. These studies agree that the use of BCI in VR games may enhance the immersion feeling. The business overview presented in [10] also outlines that there is a concrete market with increasing demands for BCI and VR technologies, in particular for the gaming industry. As it had not been established whether the use of an HMD impacts the quality of the EEG data by interfering with the signal, in a previous study we compared the power spectrum of the EEG recorded with and without an HMD [20]. Our results showed that the quality of EEG signal is similar under the two conditions. Previous studies [16], [21]–[24] conclude that performance of a VR BCI is equal to or better than that of a personal computer (PC) BCI. Recently, study [25] has shown significant improvement in BCI training when using a modern HMD and the motor imagery paradigm. However, it is difficult to compare the results of these studies because they use different VR devices and EEG paradigms. For example, [26] describes a system consisting of an HMD with an eye-tracker, whereas [16] presents an immersive game with a CAVE.[1] Moreover, [16], [21], [24], [26], [27] describe BCI applications based on P300; [21], [23], [25], [28] applications with motor imagery; and [16], [21], [29] BCI systems based on SSVEP. References [24], [26] are studies combining an HMD and the P300 paradigm, as in the present study. However, both these studies use expensive materials and require a standalone workstation, thus their systems are not suitable for popularizing BCI+VR technology due to their price and bulk.

In contrast to these studies, the P300-based HMD user interface we developed runs independently on the smartphone and not on a PC. We also implement a robust BCI based on Riemannian geometry, meeting the functional requirements for BCIs of [15]. Finally, we correct the tagging latency in VR and PC, which has never been done before, although it must be corrected to compare the ERPs in the two conditions.

The remainder of this article is organized as follows: Section II describes the materials and methods. Three kinds of analysis are presented: analysis of the EEG data, comparison of the BCI performance in VR versus PC and analysis of the user experience through a questionnaire. The results are discussed in Section III. Section IV presents our conclusion.

## II. MATERIALS AND METHODS

### A. Participants

A total of 21 volunteers participated in the experiment (7 females), with mean (sd) age 26.38 (5.78) and median age 26 years. Eighteen of the subjects were between 19 and 28 years old. The three subjects outside this range were 33, 38 and 44 years old. Before the experiment, each subject was informed that he or she would be exposed to electromagnetic radiation, as the device contained an active smartphone placed in front of the eyes. We excluded from the study all participants presenting with a risk of epilepsy or reporting previous experience with motion sickness. All participants provided written informed consent confirming they were notified of the experimental process, the data management procedures and the right to withdraw from the experiment at any time. The study was approved by the Ethical Committee of the University of Grenoble Alpes (Comité d'Ethique pour la Recherche Non-Interventionnelle).

### B. Hardware

A VR system for the general public should be affordable and lightweight but at the same time should provide high-quality immersion and graphics. It should also be able to detect precisely the user's head position and rotation, so as to enhance the immersion feeling, as well as to minimize the inter-oculus latency to improve the detection of the P300 signal. For this study, we chose a passive HMD mask manufactured by VRElegiant (Elegiant, Austin, US) (Figure 1a) and a Huawei mate 7 (Huawei, Shenzhen, China) smartphone. The VRElegiant headset (Elegiant, Austin, US) is affordable, comfortable and adapts to a wide range of smartphones. At the time of this study, the Huawei mate 7 was a middle-range smartphone, affordable for the general public. It also has a large screen (1920 x 1080), which is a desirable property to improve the immersion feeling in VR. In addition, it has a low inter-oculus latency in comparison to, for example, the Samsung S6 (Samsung, Seoul, South Korea). These considerations and the others that led to the choice of the VR material we used are detailed in a separate technical report [30].

In the PC condition, the application was run by a mid-range laptop. We found that the use of a standard i5 processor from Intel (Santa Clara, US) with an integrated graphic chipset could run our application without problems. Still, the laptop may use up to 16 GB of RAM, while RAM use on the smartphone is restricted to 2 GB. This and other considerations are to be taken into account if the data acquisition and processing are deployed to the smartphone. The screen of the laptop was a standard LCD screen with a refresh rate of 60 Hz and a resolution of 1920 x 1080 pixels. The displayed texture looked the same as that seen on the smartphone screen except for some momentary effects of pixelation in VR.

---

[1] A CAVE is an immersive VR environment reproduced by means of projections on between three and six of the walls of a room-sized cube.

Concerning the EEG system, research is currently ongoing to provide affordable hardware with low encumbrance without sacrificing the quality of the signal. Current low-cost EEG headsets such as Emotiv (Sydney, Australia) provide an EEG signal of lower quality than that of medical or research-grade EEG equipment [31], [32]. For this reason, in this study EEG signals were acquired by means of a standard research grade amplifier (g.USBamp, g.tec, Schiedlberg, Austria) and the EC20 cap equipped with 16 wet electrodes (EasyCap, Herrsching am Ammersee, Germany), placed according to the 10-20 international system. The locations of the electrodes were FP1, FP2, FC5, FC6, FZ, T7, CZ, T8, P7, P3, PZ, P4, P8, O1, Oz and O2. The reference was placed on the right earlobe and the ground at the AFZ scalp location. The amplifier was linked by USB connection to the PC, where the data were acquired by means of the open-source software OpenVibe [33], [34]. Data were acquired with no digital filter applied and a sampling frequency of 512 samples per second. For the ensuing analysis, tags were sent by the application to the amplifier through the USB port of the PC or smartphone. They were then recorded along with the EEG signal as a supplementary channel. The tagging process was the same on PC and VR with two exceptions: for the smartphone (VR), a mini-USB to USB adapter was necessary and different serial port communication libraries were used for the VR and PC.[2]

*C. Procedures*

For all subjects, the experiment took place in a small room containing the laptop, the VR headset, the smartphone and all the required hardware materials for acquiring the EEG data. Subjects sat in front of the laptop. They were instructed to avoid movement and to keep the same position during the whole experiment. These instructions were the same for all the experimental conditions. The two experimental setups are depicted in Figure 2.

To compare the use of BCI with an HMD (VR) and without an HMD (PC), we developed a simple P300 interface consisting of a six-by-six matrix of white flashing crosses. The task of the subjects was to focus on a red square target (Figure 3). The user interface was identical for the PC and VR conditions. It was implemented within the Unity engine (Unity, San Francisco, US) before being exported to the PC and VR platforms. In this way, we ensured that the visual stimulations were identical in the two experimental conditions.

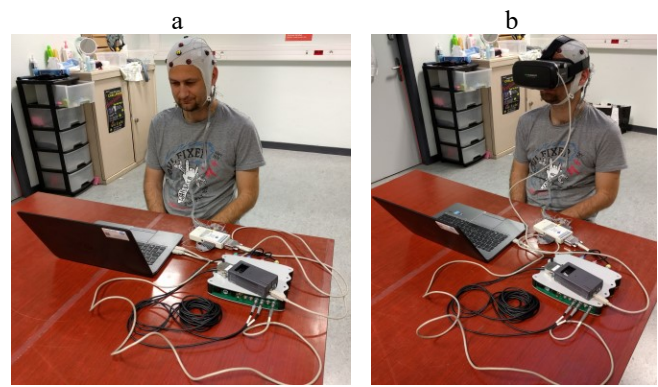

Figure 2. Experimental setup in condition PC (a) and VR (b).

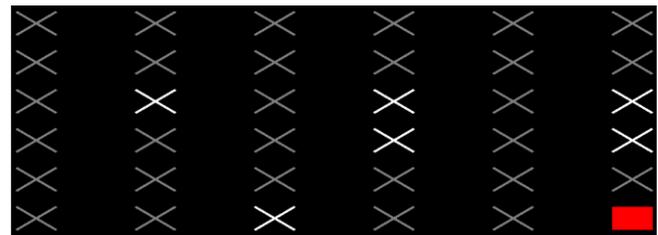

Figure 3. User interface at the moment when a group of six nontarget symbols (crosses) are flashing (in white).

The experiment was composed of two sessions. One session ran under the PC condition and the other under the VR condition. The order of the sessions was randomized for all subjects. Each session comprised 12 blocks of five repetitions (Figure 4). A repetition consisted of 12 flashes of groups of six symbols chosen in such a way that over the course of one repetition, each of the 36 symbols flashed exactly two times [35], [36]. Thus, in each repetition, two groups of six flashing symbols included the target, whereas the remaining 10 flashes where composed of a group of six nontarget symbols. The target symbol was the same for all five repetitions within a block. That is, a target symbol flashed exactly 10 times within a block (5 repetitions x 2 target flashes). The onset of each flash was tagged into the EEG stream.

After each block of five repetitions with the same target, a random feedback message was given to the subject in the form of the item selection. A 2s pause was allowed between the end of the repetition and the release of the feedback in order to mimic network latency (such as between the smartphone and OpenVibe). The feedback was "correct" if the selected symbol was the target, "incorrect" otherwise. The feedback was drawn randomly from a uniform distribution with "correct" representing 70% of results. The use of random feedback ensures that the performance of a participant does not depend on the feedback, avoiding confounding effects due to inter-subject variability—for instance, the subject's perceived confidence or frustration in operating the BCI, which may affect his or her actual performance and concentration. At the end of the experiment, the user answered a questionnaire, reported at the end of this document.

---

[2] https://github.com/mik3y/usb-serial-for-android (smartphone)   https://github.com/manashmndl/SerialPort (PC)

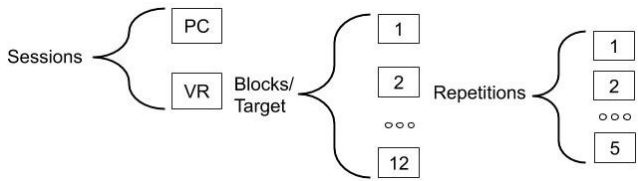

Figure 4. The experiment for each participant was composed of two sessions of 12 blocks, a block consisting of five repetitions of the same target.

A pilot experiment showed that the inertial measurement unit (IMU) of the smartphone sometimes accumulated an unexpected amount of drift, causing the virtual world to slowly move around the subject. Therefore, the IMU was deactivated for the experiments. As a consequence, the application was always fixed in front of the subject's eyes.

### D. EEG data analysis

#### 1) Method

As pre-processing, we applied a fourth-order linear phase response IIR (Infinite Impulse Response) Butterworth filter in the bandpass region 1–20 Hz. Then we used a linear phase response IIR notch filter at 50Hz with a Q factor equal to 35. These filters were implemented in Matlab (Mathworks, Natick, USA) using the *butter* and *iirnotch* functions associated with the *filtfilt* function, which implements zero-phase digital filtering. The data were then down-sampled to 128 samples per second. The ensuing analysis was carried out using in-house software and the Brainstorm software [37]. We extracted from the signal epochs of 1s of EEG data after each tag. The timestamps of the tags were corrected by taking into account the average latency of the tagging in the two conditions. The latency of the tagging method was measured for both the left and right screens in the VR condition. We kept the smaller of these two measures, as it corresponded to the first appearance of the stimulus on the screen [38]. All ERP epochs were shifted with respect to the tag according to the latency estimation. The estimated latencies (sd) for PC and VR were 38.1 (5.3) ms and 117.23 (5.81) ms, respectively. There were a total of 120 target epochs (12 blocks x 5 repetitions x 2 flashes) and 600 nontarget epochs (12 blocks x 5 repetitions x 10 flashes) per subject for each experimental condition. Each set of 120 target epochs and 600 nontarget epochs were arithmetically averaged. Then, we computed for each subject the difference between the average ERPs for target and nontarget epochs. These 42 average differences of ERPs (21 subjects, two conditions) were entered into a paired, two-sided, cluster-based permutation test [39] comparing the VR versus the PC condition (Figure 5). In our case, clusters were constituted on the basis of the temporal (EEG samples from 0 to 1 s post-stimulus) and spatial (all scalp electrodes) adjacency of the effect. The cluster-based approach allows circumventing of the multiple comparison problem— i.e., it ensures that the type I error rate is below the predefined alpha level, which in this study we set to the typical 0.05 level. The test was run using the *ft_timelockstatistics* routine in Fieldtrip [40], [41] (included in Brainstorm). In this routine, the *cluster alpha threshold* was set to 0.025. An approximate p-value was obtained by means of 5,000 random permutations.

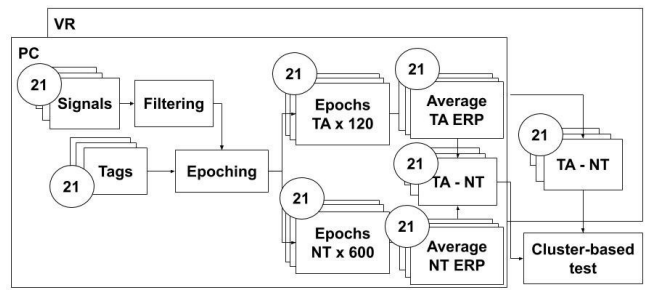

Figure 5. Flowchart of the data analysis. TA and NT denote target and nontarget, respectively.

#### 2) Results

In both conditions, ERPs of interest are found between about 100 ms and 700 ms (Figure 6). The cluster-based test revealed significant differences between the two conditions in the 148–313-ms range in the central, frontal, left temporal, parietal and occipital locations (differences were not significant only for electrodes FP1, FP2, FC6 and T8). This time interval corresponds to the P200, which is statistically wider and larger in VR as compared to PC.

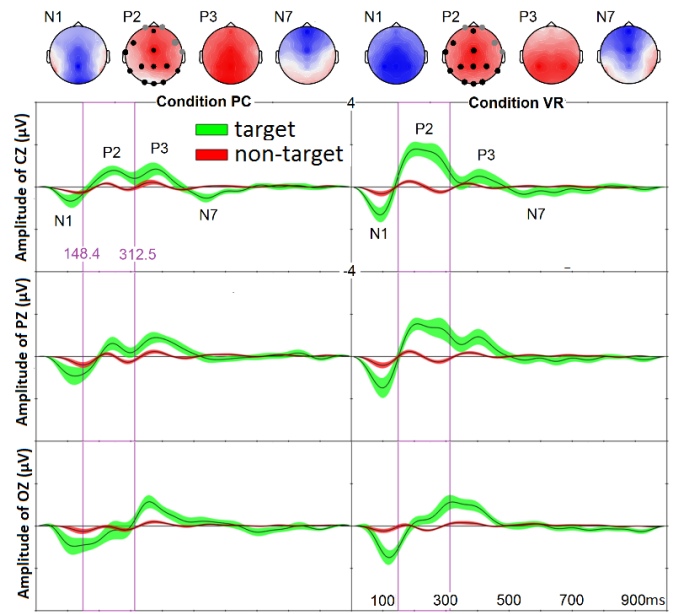

Figure 6. From zero to one second after stimulation: grand average (21 subjects) of the signal at the CZ, PZ and OZ electrodes (thick lines). The colored areas in orange and green display the ±1standard error areas of the nontarget and target ERP, respectively. At top are the scalp topographies of the grand average of the amplitude of the target minus nontarget epochs, averaged from 10 ms before to 10 ms after each peak. The two vertical lines enclose the time period where the permutation test detected a significant difference between the PC and the VR conditions. The electrodes marked by black disks comprise the significant cluster.

Qualitative but not significant differences also appear in the shape, amplitude and latency of the other ERP components, which however all have similar topographies in the two conditions:
- The peak of the early negative visual potential (N100) with occipital topographic dominance appears ~15 ms later and with lower amplitude in the PC condition as compared to VR.

- The P300 peak occurs with maximal amplitude at around 380 ms in the PC condition and at around 400 ms in the VR condition, in both cases with central and occipital dominance but with lower amplitude in the VR condition.
- A late negativity (probably an N700) with frontal and central dominance appears with maximal amplitude at around 540 ms in both PC and VR, with a higher amplitude in PC. The negativity begins earlier in PC (480 ms vs 500 ms in VR) but perdures longer in VR (~200 ms in PC vs 280 ms in VR).

While we show here only the traces at representative electrodes CZ, PZ and OZ, the same peaks are identifiable in most of the electrodes.

### E. Comparison of BCI performance

#### 1) Method

In this section, we test the performance of an *offline* classifier in the two experimental conditions. We extracted epochs of 600 ms after each tag. We applied a simplified version of the spatial filtering described in [42] to improve the signal-to-noise-ratio. In summary, let us denote $TA$ the set of size $K$ containing all the targets epochs and $NT$ the set of size $L$ containing all the nontarget epochs. We compute $C$, the mean of the covariance matrices of all epochs, as

$$C = \frac{1}{K+L}\sum_{X \in (TA \cup NT)} XX^T$$

and compute $C_{TA}$, the covariance matrix of the evoked potential of the target epochs, as

$$C_{TA} = \bar{X}_{TA}\bar{X}_{TA}^T, \quad \text{where} \quad \bar{X}_{TA} = \frac{1}{K}\sum_{X \in TA} X.$$

We then compute the generalized eigenvalue decomposition of $C$ and $C_{TA}$ as

$$UCU^t = I_n \text{ and } U C_{TA} U^t = \Lambda,$$

where $U$ is an invertible matrix, and $\Lambda$ is diagonal and holds the generalized eigenvalues. Notice that the elements of $\Lambda$ are also the eigenvalues of $C^{-1}C_{TA}$. Spatial filtering implies a dimensionality reduction. To define the subspace, we take the four generalized eigenvectors of $U$ corresponding to the four largest eigenvalues in $\Lambda$. These eigenvectors correspond to the components that maximize the ratio between $C_{TA}$ and $C$ and thus are the most discriminative.

For cross-validation purposes, the 12 blocks were separated into training and testing sets. To determinate the optimum number of blocks for training, the performance values of the classifier were assessed for different training sizes, ranging from 10% to 90% of the total number of blocks in steps of 10%. For each training set and condition (PC or VR), we randomly selected epochs from the training blocks to build a Riemannian minimum-distance-to-mean (RMDM) classifier [15], [43] and used the remaining blocks for testing. We implemented the RMDM algorithm using the log-determinant distance and mean [44].

For each training set, we tested the RMDM classifier for different numbers of repetitions, from one to five, for each test block. When using more than one repetition, we averaged together the epochs obtained in each repetition in the two flashing conditions (target and nontarget). The performance was assessed using three metrics. The first metric is the hit rate (HR), which is the proportion of time the target is correctly identified by the classifier. Metric HR is interesting from the user perspective, since it naturally reflects the performance of the user according to the task. HR is also useful to compute the information transfer rate (ITR), which is a standard measure to evaluate the responsiveness of a BCI [3] in bit/min. It is defined by $\frac{log2(N)+Plog2(P)+(1-P)log2\frac{1-P}{N-1}}{T}$, where $N$ is the number of symbols, $P$ the accuracy of the selection and $T$ the average time to select a symbol. The second metric is the balanced accuracy (BA), which is defined by $\frac{1}{2}(\frac{A}{A+B} + \frac{C}{C+D})$, where $A$ and $B$ (respectively $C$ and $D$) stand for the number of correctly and incorrectly classified flashes of nontarget (resp. target) groups. In comparison to the HR, the BA takes into account also the rate of correctly classified nontarget symbols. The third metric is the area under the receive operating characteristic curve (ROC-AUC), which is a standard measure to evaluate the performance of a classifier for unbalanced classes, although the resulting score is less intuitive. Unlike the HR metric, but like the BA metric, ROC-AUC is a flash-based metric and not a repetition-based metric. The area under the curve (AUC) of each metric was computed to provide a unique index of the performance of each training set. Metrics were averaged over 100 randomly chosen sets (Figure 7). As shown in Figure 8, the AUC of the classifier displays a logarithmic profile with a plateau starting at about 40 epochs. Since the AUC does not improve much after 40 epochs, we kept this figure for training size. This implies that the optimum size of the training set is around 30% (40/120 epochs) of the total number of blocks, independently of the experimental condition. We conclude that the VR condition requires the same amount of data for training as the PC condition.

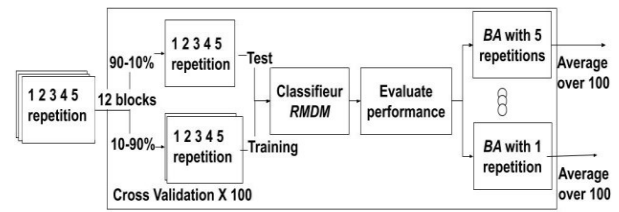

Figure 7. System flowchart of the method for comparison of BCI performance.

The difference in the mean classification accuracy between the two conditions as a function of the number of repetitions (one to five) was evaluated using the BA metric by means of a two-way within-subject analysis of variance (ANOVA) where the first factor was the experimental condition and the second factor was the number of repetitions. Only one metric (BA) was used to keep a reasonable level for the type I errors (fixed to α = 0.05). The ANOVA tests were conducted with and without the three outsiders (i.e., the three subjects older than 28 years). There was no difference in the analysis with and without outsiders, thus we present the results obtained with all subjects.

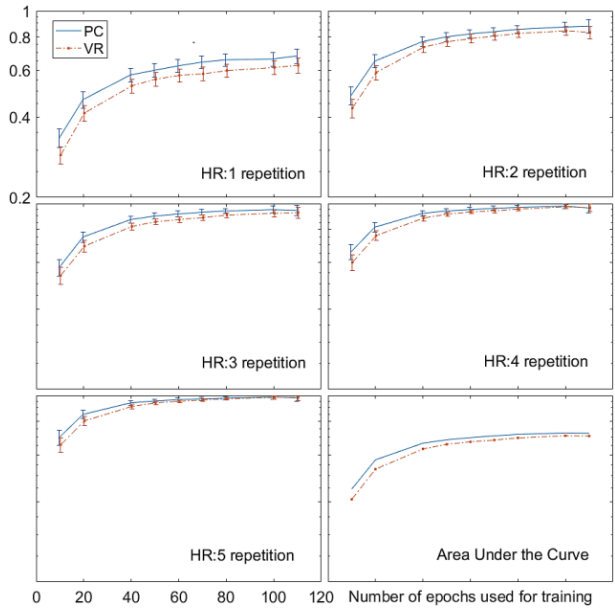

Figure 8. Grand average of balanced accuracy according to the number of epochs used for training. The vertical lines are the standard errors: that is, the standard deviations divided by the square root of the number of subjects.

*2) Results*

In Figure 9, the HR for PC (VR) is in the range 0.57–0.94 (0.52–0.91) depending on the number of repetitions. Although maximum accuracy selection was obtained with five repetitions, the ITR is maximal with one repetition, as the required time to select a symbol increases as a function of the repetition number. The highest ITR was 20.1 bit/min for PC and 17.31 bit/min for VR.

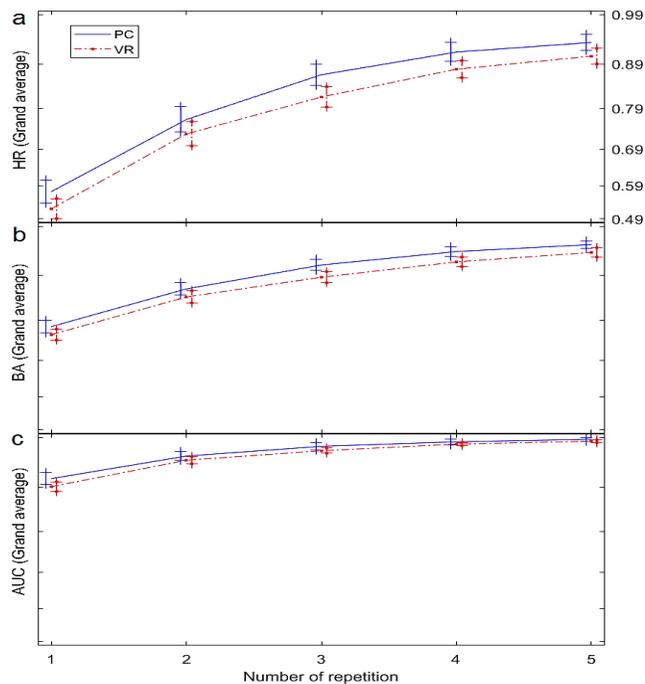

Figure 9. Grand average of the classification accuracy obtained with the three performance metrics as a function of the number of repetitions. Metrics: hit rate (HR) (a), balanced accuracy (BA) (b) and ROC-AUC (c). 60% of the blocks were used for training, totaling 70 target epochs. The vertical lines are the standard errors.

The ANOVA revealed no difference in classification accuracy obtained in the two experimental conditions. There was a main effect on the repetition factor ($p < 0.001$). This effect is well expected, since averaging evoked potentials across repetitions increases the signal-to-noise ratio and improves the classification accuracy (Figure 9).

### F. Analysis of the questionnaire

*1) Method*

At the end of the experimental session, the subjects answered a questionnaire, reported at the end of this document. From the questionnaire, we extracted four variables (Table 2): *gender* (Male or Female), *amount of previous experience in VR* (none, occasional or repetitive experience), *sensation of discomfort* and *sensation of control preference* (SCP). SCP is a subjective index taking one of two possible values depending on whether the sensation of control is greater in VR or in PC. The differences between the means observed in the levels of these four variables were assessed by means of four two-way mixed-model ANOVAs, where the above variables were the between-subject factor and the experimental condition (PC or VR) was the within-subject factor. Using the method of Bonferroni, we adjusted the p-values threshold to $\alpha = 0.0125$ (*i.e.,* 0.05/4).

Additionally, we tested whether a better sensation of control in VR (PC) was correlated to a preference for VR (PC) using the phi coefficient [45], which measures the association of two binary variables (here the SCP and the preference for VR or PC, which was one if the subject preferred VR and zero otherwise).

*2) Results*

Descriptive statistics of interest are as follows:

- 66.67% (2.16) of the subjects preferred the VR condition.
- 61.90% (2.23) of the subjects had a better or equal sensation of control under the VR condition as compared to PC.
- The experiment resulted in fatigue for 52.38% (2.29) of the subjects. This is consistent with previous studies reporting a high mental load with the visual P300 (*e.g.,* [46]). This outlines the need for smart design when using BCI with VR [10].
- The mean (sd) sensation of control values for VR and PC were 6.48 (2.73) and 6.33 (2.58) on a 10-point scale which is in line with the probability of receiving correct feedback (70%).
- 28.57% (2.07) of the subjects reported a sensation of discomfort due to a heavy HMD, the effort of concentration, visual fatigue, a problem setting up the HMD or the flashing of stimuli groups distracting them from the target. These problems were also reported by subjects who did not report a sensation of discomfort.

Note that for binary variables, the standard deviations indicated within the brackets were computed using a binomial law. Participants also provided general remarks, summarized as follows:

- Include the correction of their vision in the HMD. Also, setting the stereo convergence takes time for many subjects.
- Add feedback for error quantification. Indeed, it was reported that when the result was incorrect, there was no indication on how close the participant has come to being correct, making it difficult to improve performance.

- Some participants reported that the low performance was expected because of their lack of focus. Others said that they found tricks to improve their accuracy by concentrating harder on the target, viewing all the crosses in the periphery or counting the flashes in their head.

None of the ANOVA test results were significant. However we mention that the type II error rate for these tests, which we estimated using the g*power software [47], [48] that follows the method of Cohen [49], was rather high.

The result from the phi-test showed a significant correlation (phi-coefficient = 0.67, n = 14, $p < 0.05$) between a higher sensation of control for a condition and a preference for that condition. This held for both conditions (phi-coefficient = 0.48, n = 7, $p < 0.05$).

To summarize, effects of discomfort were reported by participants whether or not they answered yes to the sensation of discomfort question on the questionnaire, suggesting that there is no link between these remarks and the sensation of discomfort. Results from the questionnaire also suggest four general remarks. First, people tried to explain their performance even though they had no control over their performance. Second, the sensation of control was in line with the given feedback. Third, the sensation of control was higher in the preferred experimental condition. Fourth, the majority of subjects preferred the BCI running in VR.

## III. Discussion

The P200 component of the ERP was significantly different in the VR and PC conditions. The P200 is part of the long latency response, occurring after sensory responses and before high-level cognitive tasks such as the P300 [50]. It is involved in the cognitive process comparing sensory inputs with memory [51] and is modulated by arousal, attention [50], depth perception [52] or the intensity of the stimuli (see [24]).

We also found that the N100 component was more pronounced in VR than in PC, although this difference was not statistically significant. Since the N100 component is produced by the two parts of the brain according to the location of the stimulus, the effect of VR stereoscopy on the N100 was expected and already documented in [53].

A late negativity appears in both conditions. It may be the N700, an ERP component involved in the cognitive task of determining concreteness [54]–[56], and may be due to the geometry of the symbols we used (square, cross).

The P300 component was very similar in the PC and VR conditions. Results from previous studies comparing stereoscopic vision with normal vision do not agree. In [57], the authors found that stereoscopy elicits higher but delayed P300, whereas [58] did not find any significant differences between the two conditions. The two studies used a polarized monitor, but with a different refresh rate (60 and 240 frames-per-second in the first and second studies, respectively). The duration of the stimuli as well as the inter-stimuli interval were also considerably longer in [58] (respectively, 500 and 500 ms versus 100 and 30 ms).

This comparison of the ERPs in PC and VR has been empowered by the correction of the timestamps of the tags using the average latency of the tagging in the two conditions. Surprisingly, it exists a noticeable difference between these two latencies (38.1 ms in PC vs 117.23 ms in VR). As we explained in [38], this difference is caused by hardware and software implementations, whereas in general is not related to the subject's capability – except for people having an uncommon perception of the display image rate such as, for instance, hardcore gamers or pilots. In [30] we exposed the methodology for computing these latencies.

The variations in ERP amplitude we found are likely due to the different size and luminosity of the stimuli in VR and PC. They can also be explained by the latency between the two parts of the screen in VR, as a high latency between the screens causes the stimulus to remain displayed longer. However, the effect of the duration of the stimulus on the ERP amplitude and classification may not be meaningful [59]. To our knowledge, no previous article has compared stereoscopic and normal vision with an HMD while correcting appropriately for the latency, therefore our findings may be considered new.

The AUC of our classifier ranges between 0.9 and 1.0, which is almost the same as the AUC reported in [31]. In spite of a higher ERP amplitude in VR, there were not significant differences in classification accuracy between the VR and PC conditions, which may be explained by the fact that the P200 component is stronger but also more variable in the VR condition (Figure 6). This result is consistent with a previous study on HMDs [24].

We chose to disable the IMU to avoid the drift. The IMU is part of the immersion process in VR. Thus by disabling it, admittedly we did not make use of the main capabilities of the VR system, which are to move in and watch in a 3D environment. For instance, we may expect that the 3D immersion has a higher impact on the sensation of control in VR, thus resulting in a significant effect on performance in VR. However, our objective was to assess whether the same BCI task performed on a PC may also be achieved in VR using inexpensive equipment, which to the best of our knowledge has never been done before. The present study answers this basic requirement, providing a first point of comparison for further studies willing to develop options for BCI+VR technology for the general public.

In this study we also chose to concentrate on hardware tagging through the USB port because it had already been tested in [35]. However, by synchronizing clocks in the HMD and the acquisition system, it would be possible to communicate without a cable. This method is enabled in the OpenVibe platform [60] and could be used in building a mobile HMD with a P300-based BCI. Indeed, the development of inside-out technology allows the use of mobile HMDs. However, the possibility to classify P300s when the subject is moving is still a subject of research [61]–[63]. The use of mixed interfaces, by taking into account eye blinks [64] or auditory stimuli [46], [65], for example, is another option to circumvent these limitations. In particular, the gyroscope and accelerometer integrated in the smartphone may participate in the recognition of the user's intention, or at least in the detection of artifacts.

This led us to the main limitation of the system we implemented, which is due to movement. While movement is a basic input for VR, it interferes with the acquisition of EEG signals. At the same time, popular VR applications such as action games require speed and accuracy, which are incompatible with the use of BCIs. Some of these aspects have been studied further in a previous publication [10]. There, we suggested restricting BCI to a few elements of the application. In general, applications with a slow gameplay are suitable, such as turn-based simulation or adventure games.

In this same vein, the analysis of the responses to the questionnaire highlights the impact of subjective factors on the user experience. In particular, we found that the subject's sensation of control is higher in his or her preferred condition (PC or VR) and that the sensation of control is in line with the given feedback, although these results need further investigation to be generalized. In practice, this suggests that guidance is a suitable way to compensate for unreliable inputs while providing a convincing sensation of control. In this line, choosing the type of device as a function of user preference might effectively enhance the sensation of control independently of the displayed application.

Taken together, the last remarks are interesting for designing gaming applications, suggesting that the design of the application is more important for gaming than the accuracy of the BCI itself. This conclusion was drawn in [18], [66], where it was suggested that unreliable input can be used to develop fun games. Reference [18] integrates a BCI control into a popular role-play fantasy game. The authors found that in spite of weak control and involvement using such BCI—in particular, the accuracy was around 75%—the experience of fun was similar with and without BCI control. Although highly unreliable controls often result in frustration, it was also shown that players had less fun while experiencing a game with perfect control [66].

## IV. CONCLUSION

The introduction of VR+BCI technology has not been investigated using affordable materials such as a passive HMD running on an ordinary smartphone. Yet, the price of high-end VR materials is an obstacle to the popularization of a mixed technology VR+BCI. This study evaluated the performance of a BCI displayed on an ordinary VR device (a smartphone) in comparison to a BCI displayed on a PC, while also assessing several user experience factors. We showed that the performance of a P300-based BCI coupled with a passive HMD is comparable to the performance of a state-of-the-art BCI displayed on a PC. In other words, proper ERPs are elicited using such an HMD, and the average classification accuracy is adequate. These results extend those of [24], which were gathered using the same kind of BCI but displayed on an expensive HMD. The data from the present study are freely available for download at https://doi.org/10.5281/zenodo.2605204.

## ACKNOWLEDGEMENT

The author would like to thank Dr. Simon Barthelme for the fruitful discussions on statistical tests used in this article, Dr. Florent Bouchard for his help on the implementation of the spatial filter and Mr. Maxime Delaporte, game designer, for the fruitful discussions on control and sensation of control in gaming.

## QUESTIONNAIRE

The questions asked of participants in the questionnaire are presented in Table 1. In the analysis, we used the factors presented in Table 2. As an inclusion criterion, we submitted to statistical analysis only the factors containing at least six participants for each level. When the question was open, such as "How many hours a week do you play first-person shooters?" the levels were created by the authors. In some cases, there were not enough participants to fulfill our criterion, as was the case for questions 1, 2, 8 and 9 (see Table 1).

| Number | Question |
|---|---|
| 1 | Evaluate your tiredness before the experiment on a scale from 0 to 10 where 0 is "no fatigue." |
| 2 | Evaluate your tiredness after the experiment on a scale from 0 to 10 where 0 is "no fatigue." |
| 3 | Did you feel a sensation of discomfort? |
| 4 | If yes, why? (free answer) |
| 5 | Did you prefer the PC or VR session (answer: PC, VR, SAME)? |
| 6 | Evaluate your sensation of control under PC on a scale from 0 to 10 (0 = "no control"). |
| 7 | Evaluate your sensation of control under VR on a scale from 0 to 10 (0 = "no control"). |
| 8 | How many hours a week do you play video games? |
| 9 | How many hours a week do you play first-player shooter? |
| 10 | Have you ever experienced virtual reality? If yes, how many times? |
| 11 | Do you have any suggestions or remarks concerning the experiment? |
| 12 | Please circle your gender: Male - Female. |

Table 1. Questionnaire

| Factor | Levels |
|---|---|
| Amount of experience in VR | According to the answers to question 10 subjects were categorized in three groups:1 for none; 2 for occasional and 3 for repeated experience in VR. |
| Sensation of control preference | 1 if the sensation of control under PC was greater than the sensation of control under VR, 3 if vice versa. |
| Gender | 1 for male, 0 for female. |
| Discomfort | 1 for a positive answer to question 3, 0 elsewhere. |

Table 2. Description of factors and their levels

## REFERENCES

[1] M. Balaji, *Thinking Dead: What the Zombie Apocalypse Means*. Lexington Books, 2013.

[2] F. Biocca, 'The Cyborg's Dilemma: Progressive Embodiment in Virtual Environments', *J. Comput.-Mediat. Commun.*, vol. 3, no. 2, Sep. 1997.

[3] J. Wolpaw and E. W. Wolpaw, *Brain-Computer Interfaces: Principles and Practice*. Oxford University Press, USA, 2012.

[4] J. J. Vidal, 'Toward Direct Brain-Computer Communication', *Annu. Rev. Biophys. Bioeng.*, vol. 2, no. 1, pp. 157–180, 1973.

[5] J. J. Vidal, 'Real-time detection of brain events in EEG', *Proc. IEEE*, vol. 65, no. 5, pp. 633–641, May 1977.

[6] L. A. Farwell and E. Donchin, 'Talking off the top of your head: toward a mental prosthesis utilizing event-related brain potentials', *Electroencephalogr. Clin. Neurophysiol.*, vol. 70, no. 6, pp. 510–523, Dec. 1988.

[7] N. Birbaumer and L. G. Cohen, 'Brain–computer interfaces: communication and restoration of movement in paralysis', *J. Physiol.*, vol. 579, no. 3, pp. 621–636, Mar. 2007.

[8] E. W. Sellers and E. Donchin, 'A P300-based brain–computer interface: Initial tests by ALS patients', *Clin. Neurophysiol.*, vol. 117, no. 3, pp. 538–548, Mar. 2006.

[9] T. O. Zander and C. Kothe, 'Towards passive brain–computer interfaces: applying brain–computer interface technology to human–machine systems in general', *J. Neural Eng.*, vol. 8, no. 2, p. 025005, 2011.

[10] G. Cattan, C. Mendoza, A. Andreev, and M. Congedo, 'Recommendations for Integrating a P300-Based Brain Computer Interface in Virtual Reality Environments for Gaming', *Computers*, vol. 7, no. 2, p. 34, May 2018.

[11] Y. Zhang, P. Xu, T. Liu, J. Hu, R. Zhang, and D. Yao, 'Multiple Frequencies Sequential Coding for SSVEP-Based Brain-Computer Interface', *PLOS ONE*, vol. 7, no. 3, p. e29519, Mar. 2012.

[12] F. Sepulveda, 'Brain-actuated Control of Robot Navigation', in *Advances in Robot Navigation*, vol. 8, Alejandra Barrera, 2011.

[13] A. Barachant and M. Congedo, 'A Plug&Play P300 BCI Using Information Geometry', *ArXiv14090107 Cs Stat*, Aug. 2014.

[14] M. Congedo, 'EEG Source Analysis', Habilitation à diriger des recherches, Université de Grenoble, 2013.

[15] M. Congedo, A. Barachant, and R. Bhatia, 'Riemannian geometry for EEG-based brain-computer interfaces; a primer and a review', *Brain-Comput. Interfaces*, vol. 4, no. 3, pp. 155–174, 2017.

[16] A. Lécuyer, F. Lotte, R. B. Reilly, R. Leeb, M. Hirose, and M. Slater, 'Brain-Computer Interfaces, Virtual Reality, and Videogames', *Computer*, vol. 41, no. 10, pp. 66–72, Oct. 2008.

[17] D. Marshall, D. Coyle, S. Wilson, and M. Callaghan, 'Games, Gameplay, and BCI: The State of the Art', *IEEE Trans. Comput. Intell. AI Games*, vol. 5, no. 2, pp. 82–99, Jun. 2013.

[18] B. van de Laar, H. Gürkök, D. P.-O. Bos, M. Poel, and A. Nijholt, 'Experiencing BCI Control in a Popular Computer Game', *IEEE Trans. Comput. Intell. AI Games*, vol. 5, no. 2, pp. 176–184, Jun. 2013.

[19] I. P. Ganin, S. L. Shishkin, and A. Y. Kaplan, 'A P300-based Brain-Computer Interface with Stimuli on Moving Objects: Four-Session Single-Trial and Triple-Trial Tests with a Game-Like Task Design', *PLOS ONE*, vol. 8, no. 10, Oct. 2013.

[20] G. Cattan, A. Andreev, C. Mendoza, and M. Congedo, 'The Impact of Passive Head-Mounted Virtual Reality Devices on the Quality of EEG Signals', in *Workshop on Virtual Reality Interaction and Physical Simulation*, Delft, 2018.

[21] F. Lotte, 'Les Interfaces Cerveau-Ordinateur: Conception et Utilisation en Réalité Virtuelle', *Rev. Sci. Technol. Inf. - Sér. TSI Tech. Sci. Inform.*, vol. 31, no. 3, pp. 289–310, 2012.

[22] B. H. Cho et al., 'Attention Enhancement System using virtual reality and EEG biofeedback', in *Proceedings IEEE Virtual Reality 2002*, 2002, pp. 156–163.

[23] R. Ron-Angevin and A. Díaz-Estrella, 'Brain-computer interface: changes in performance using virtual reality techniques', *Neurosci. Lett.*, vol. 449, no. 2, pp. 123–127, Jan. 2009.

[24] I. Käthner, A. Kübler, and S. Halder, 'Rapid P300 brain-computer interface communication with a head-mounted display', *Front. Neurosci.*, vol. 9, p. 207, 2015.

[25] F. Škola and F. Liarokapis, 'Embodied VR environment facilitates motor imagery brain–computer interface training', *Comput. Graph.*, Jun. 2018.

[26] J. D. Bayliss and D. H. Ballard, 'A virtual reality testbed for brain-computer interface research', *IEEE Trans. Rehabil. Eng. Publ. IEEE Eng. Med. Biol. Soc.*, vol. 8, no. 2, pp. 188–190, Jun. 2000.

[27] J. D. Bayliss, 'Use of the evoked potential P3 component for control in a virtual apartment', *IEEE Trans. Neural Syst. Rehabil. Eng.*, vol. 11, no. 2, pp. 113–116, Jun. 2003.

[28] F. Lotte, Y. Renard, and A. Lécuyer, 'Self-Paced Brain-Computer Interaction with Virtual Worlds: A Quantitative and Qualitative Study "Out of the Lab"', in *4th international Brain Computer Interface Workshop and Training Course*, Graz, Austria, 2008.

[29] J. Legény, R. V. Abad, and A. Lécuyer, 'Navigating in Virtual Worlds Using a Self-Paced SSVEP-Based Brain #8211;Computer Interface with Integrated Stimulation and Real-Time Feedback', *Presence*, vol. 20, no. 6, pp. 529–544, Dec. 2011.

[30] A. Andreev, G. Cattan, and M. Congedo, 'Engineering study on the use of Head-Mounted display for Brain-Computer Interface', GIPSA-lab, Technical Report 1, Jun. 2019.

[31] L. Mayaud, M. Congedo, A. Van Laghenhove, M. Figère, E. Azabou, and F. Cheliout-Heraut, 'A comparison of recording modalities of P300 event-related potentials (ERP) for brain-computer interface (BCI) paradigm', *Neurophysiol. Clin. Neurophysiol.*, vol. 43, no. 4, pp. 217–227, Oct. 2013.

[32] D. M. Buchanan, J. Grant, and A. D'Angiulli, 'Commercial wireless versus standard stationary EEG systems for personalized emotional brain-computer interfaces: a preliminary reliability check', *Neurosci. Res. Notes*, vol. 2, no. 1, pp. 7–15, Mar. 2019.

[33] Y. Renard et al., 'OpenViBE: An Open-Source Software Platform to Design, Test, and Use Brain–Computer Interfaces in Real and Virtual Environments', *Presence Teleoperators Virtual Environ.*, vol. 19, no. 1, pp. 35–53, Feb. 2010.

[34] C. Arrouët, M. Congedo, J.-E. Marvie, F. Lamarche, A. Lécuyer, and B. Arnaldi, 'Open-ViBE: A Three

Dimensional Platform for Real-Time Neuroscience', *J. Neurother.*, vol. 9, no. 1, pp. 3–25, Jul. 2005.

[35] A. Andreev, A. Barachant, F. Lotte, and M. Congedo, *Recreational Applications of OpenViBE: Brain Invaders and Use-the-Force*, vol. chap. 14. John Wiley ; Sons, 2016.

[36] M. Congedo *et al.*, '"Brain Invaders": a prototype of an open-source P300- based video game working with the OpenViBE platform', in *5th International Brain-Computer Interface Conference 2011 (BCI 2011)*, 2011, pp. 280–283.

[37] F. Tadel, S. Baillet, J. C. Mosher, D. Pantazis, and R. M. Leahy, 'Brainstorm: A User-Friendly Application for MEG/EEG Analysis', *Computational Intelligence and Neuroscience*, 2011. [Online]. Available: https://www.hindawi.com/journals/cin/2011/879716/.

[38] G. Cattan, A. Andreev, B. Maureille, and M. Congedo, 'Analysis of tagging latency when comparing event-related potentials', Gipsa-Lab ; IHMTEK, Grenoble, Technical Report, Dec. 2018.

[39] T. E. Nichols and A. P. Holmes, 'Nonparametric permutation tests for functional neuroimaging: a primer with examples', *Hum. Brain Mapp.*, vol. 15, no. 1, pp. 1–25, Jan. 2002.

[40] R. Oostenveld, P. Fries, E. Maris, and J.-M. Schoffelen, 'FieldTrip: Open Source Software for Advanced Analysis of MEG, EEG, and Invasive Electrophysiological Data', *Comput. Intell. Neurosci.*, vol. 2011, p. e156869, Dec. 2010.

[41] E. Maris and R. Oostenveld, 'Nonparametric statistical testing of EEG- and MEG-data', *J. Neurosci. Methods*, vol. 164, no. 1, pp. 177–190, Aug. 2007.

[42] M. Congedo, L. Korczowski, A. Delorme, and F. Lopes Da Silva, 'Spatio-temporal common pattern: A companion method for ERP analysis in the time domain', *J. Neurosci. Methods*, vol. 267, pp. 74–88, 2016.

[43] A. Barachant, S. Bonnet, M. Congedo, and C. Jutten, 'Multiclass brain-computer interface classification by Riemannian geometry', *IEEE Trans. Biomed. Eng.*, vol. 59, no. 4, pp. 920–928, Apr. 2012.

[44] Z. Chebbi and M. Moakher, 'Means of Hermitian positive-definite matrices based on the log-determinant α-divergence function', *Linear Algebra Its Appl.*, vol. 436, no. 7, pp. 1872–1889, Apr. 2012.

[45] 'Mathematical Methods of Statistics (PMS-9), Volume 9', *Princeton University Press*. [Online]. Available: https://press.princeton.edu/titles/391.html. [Accessed: 08-Dec-2017].

[46] X. An, J. Höhne, D. Ming, and B. Blankertz, 'Exploring Combinations of Auditory and Visual Stimuli for Gaze-Independent Brain-Computer Interfaces', *PLoS ONE*, vol. 9, no. 10, Oct. 2014.

[47] F. Faul, E. Erdfelder, A.-G. Lang, and A. Buchner, 'G*Power 3: a flexible statistical power analysis program for the social, behavioral, and biomedical sciences', *Behav. Res. Methods*, vol. 39, no. 2, pp. 175–191, May 2007.

[48] F. Faul, E. Erdfelder, A. Buchner, and A.-G. Lang, 'Statistical power analyses using G*Power 3.1: tests for correlation and regression analyses', *Behav. Res. Methods*, vol. 41, no. 4, pp. 1149–1160, Nov. 2009.

[49] J. Cohen, 'A power primer', *Psychol. Bull.*, vol. 112, no. 1, pp. 155–159, Jul. 1992.

[50] S. J. Luck, 'Event-related potentials', in *APA handbook of research methods in psychology, Vol 1: Foundations, planning, measures, and psychometrics*, Washington, DC, US: American Psychological Association, 2012, pp. 523–546.

[51] R. Freunberger, W. Klimesch, M. Doppelmayr, and Y. Höller, 'Visual P2 component is related to theta phase-locking', *Neurosci. Lett.*, vol. 426, no. 3, pp. 181–186, Oct. 2007.

[52] S. Omoto *et al.*, 'P1 and P2 components of human visual evoked potentials are modulated by depth perception of 3-dimensional images', *Clin. Neurophysiol. Off. J. Int. Fed. Clin. Neurophysiol.*, vol. 121, no. 3, pp. 386–391, Mar. 2010.

[53] A. J. Pegna, A. Darque, M. V. Roberts, and E. C. Leek, 'Effects of stereoscopic disparity on early ERP components during classification of three-dimensional objects', *Q. J. Exp. Psychol. 2006*, vol. 71, no. 6, pp. 1419–1430, Jun. 2018.

[54] W. C. West and P. J. Holcomb, 'Imaginal, Semantic, and Surface-Level Processing of Concrete and Abstract Words: An Electrophysiological Investigation', *J. Cogn. Neurosci.*, vol. 12, no. 6, pp. 1024–1037, Nov. 2000.

[55] H.-W. Huang, C.-L. Lee, and K. D. Federmeier, 'Imagine that! ERPs provide evidence for distinct hemispheric contributions to the processing of concrete and abstract concepts', *NeuroImage*, vol. 49, no. 1, pp. 1116–1123, Jan. 2010.

[56] J. R. Binder, C. F. Westbury, K. A. McKiernan, E. T. Possing, and D. A. Medler, 'Distinct Brain Systems for Processing Concrete and Abstract Concepts', *J. Cogn. Neurosci.*, vol. 17, no. 6, pp. 905–917, Jun. 2005.

[57] J. Qu *et al.*, 'A Novel Three-Dimensional P300 Speller Based on Stereo Visual Stimuli', *IEEE Trans. Hum.-Mach. Syst.*, vol. 48, no. 4, pp. 392–399, Aug. 2018.

[58] H. U. Amin, A. S. Malik, W. Mumtaz, N. Badruddin, and N. Kamel, 'Evaluation of passive polarized stereoscopic 3D display for visual & mental fatigues', in *2015 37th Annual International Conference of the IEEE Engineering in Medicine and Biology Society (EMBC)*, 2015, pp. 7590–7593.

[59] J. Lu, W. Speier, X. Hu, and N. Pouratian, 'The Effects of Stimulus Timing Features on P300 Speller Performance', *Clin. Neurophysiol. Off. J. Int. Fed. Clin. Neurophysiol.*, vol. 124, no. 2, pp. 306–314, Feb. 2013.

[60] N. Foy, 'TCP Tagging (Software Tagging)', *OpenViBE*, 20-May-2016. .

[61] T. M. Lau, J. T. Gwin, K. G. McDowell, and D. P. Ferris, 'Weighted phase lag index stability as an artifact resistant measure to detect cognitive EEG activity during locomotion', *J. Neuroengineering Rehabil.*, vol. 9, p. 47, Jul. 2012.

[62] G. Gargiulo, P. Bifulco, R. A. Calvo, M. Cesarelli, C. Jin, and A. van Schaik, 'A mobile EEG system with dry electrodes', in *2008 IEEE Biomedical Circuits and Systems Conference*, 2008, pp. 273–276.

[63] M. D. Vos, M. Kroesen, R. Emkes, and S. Debener, 'P300 speller BCI with a mobile EEG system: comparison to a traditional amplifier', *J. Neural Eng.*, vol. 11, no. 3, p. 036008, 2014.

[64] J. Ma, Y. Zhang, A. Cichocki, and F. Matsuno, 'A novel EOG/EEG hybrid human-machine interface adopting eye movements and ERPs: application to robot control', *IEEE Trans. Biomed. Eng.*, vol. 62, no. 3, pp. 876–889, Mar. 2015.

[65] G. Cattan, A. Andreev, C. Mendoza, and M. Congedo, 'Report on Auditory Stimulation in Brain-Computer Interfaces', Gipsa-lab ; IHMTEK, Research Report, Jan. 2019.

[66] B. van de Laar, D. P. Bos, B. Reuderink, M. Poel, and A. Nijholt, 'How Much Control Is Enough? Influence of Unreliable Input on User Experience', *IEEE Trans. Cybern.*, vol. 43, no. 6, pp. 1584–1592, Dec. 2013.